\documentclass[manuscript]{emulateapj}
\usepackage{amsmath}
\usepackage{amssymb}
\usepackage{natbib}
\usepackage{hyperref}
\usepackage{breakurl}
\usepackage{xcolor}
\usepackage{enumitem}
\setlist{leftmargin=5.5mm}
\usepackage{lineno}

\newcommand{\mail}{lilirayhk@phys.ncku.edu.tw}
\newcommand{\flux}{\,erg\,cm$^{-2}$\,s$^{-1}$}
\newcommand{\lum}{\,erg\,s$^{-1}$}

\newcommand{\cm}{\,cm$^{-2}$}
\newcommand{\nh}{$N_\mathrm{H}$}

\newcommand{\src}{V549~Vel}
\shorttitle{\textit{Fermi}-LAT Observations of \src\ 2017}
\shortauthors{Li et al.}

\begin{document}
\title{\textit{Fermi}-LAT Observations of \src\ 2017: a Sub-Luminous Gamma-Ray Nova?}
\author{
Kwan-Lok Li\altaffilmark{1,2},
Franz-Josef Hambsch\altaffilmark{3},
Ulisse Munari\altaffilmark{4},
Brian D. Metzger\altaffilmark{5},
Laura Chomiuk\altaffilmark{6},
Andrea Frigo\altaffilmark{3},
and Jay Strader\altaffilmark{6}
}

\altaffiltext{1}{Department of Physics, National Cheng Kung University, 70101 Tainan, Taiwan; \href{mailto:\mail}{\mail} (KLL)}
\altaffiltext{2}{Institute of Astronomy, National Tsing Hua University, Hsinchu 30013, Taiwan}
\altaffiltext{3}{ANS Collaboration, c/o Astronomical Observatory, 36012 Asiago (VI), Italy}
\altaffiltext{4}{National Institute of Astrophysics (INAF), Astronomical Observatory of Padova, 36012 Asiago (VI), Italy}
\altaffiltext{5}{Department of Physics, Columbia University, New York, NY 10027, USA}
\altaffiltext{6}{Center for Data Intensive and Time Domain Astronomy, Department of Physics and Astronomy, Michigan State University, East Lansing, MI 48824, USA}

\begin{abstract}
We report on the $Fermi$-LAT detection (with $\approx5.7\,\sigma$ significance) as well as the multi-wavelength analysis of the 2017 nova eruption \src. Unlike the recent shock-powered novae ASASSN-16ma and V906~Car, the optical and $\gamma$-ray light curves of \src\ show no correlation, likely implying relatively weak shocks in the eruption.
\textit{Gaia} detected a candidate progenitor of \src\ and found a parallax measurement of $\varpi=1.91\pm0.39$~mas, equivalent to a mode distance of $d\approx560$~pc (90\% credible interval of 380--1050~pc). The progenitor was also observed by the 2MASS and WISE surveys. When adopting the \textit{Gaia} distance, the spectral energy distribution of the progenitor is close to that of a G-type star. \textit{Swift}/XRT detected the supersoft X-ray emission of the nova ($kT=30$--40~keV) since day 236, and the inferred blackbody size is comparable to that of other novae assuming $d\approx560$~pc (i.e., $R_{\rm bb}\sim5\times10^8$~cm). However, there is also an unknown astrometric excess noise of $\epsilon_i=3.2$~mas found in the \textit{Gaia} data, and the inferred distance becomes controversial. If the \textit{Gaia} distance is accurate, the $\gamma$-ray luminosity of \src\ will be as low as $L_\gamma\sim4\times10^{33}$\lum, making it the least luminous $\gamma$-ray nova known so far. This may imply that the shock properties responsible for the $\gamma$-ray emission in \src\ are different from those of the more luminous events. If the nova is located farther away, it is likely a symbiotic system with a giant companion as the observed progenitor.\\
\end{abstract}

\section{Introduction}
\label{sec:intro}

\src\ (also known as ASASSN-17mt) is a 11.3-mag optical transient in the Galactic plane ($b\approx-2.3\degr$), discovered by the \textit{All Sky Automated Survey for SuperNovae} (ASAS-SN; \citealt{2014ApJ...788...48S}) on 23 September 2017 (MJD~58019; \citealt{2017ATel10772....1S}). It was first detected when the field just became observable to ASAS-SN after an $\sim80$-day seasonal gap \citep{2017ATel10772....1S}. \textit{Gaia} detected it independently as Gaia19dff\footnote{\url{http://gsaweb.ast.cam.ac.uk/alerts/alert/Gaia19dff/}} and found that it is still in quiescence on 2 September 2017, shortening the gap to 21 days.
The transient was later spectroscopically identified as a classical nova in the optically thick stage (or the ``Fe curtain'' state; \citealt{2017ATel10795....1L}). Furthermore, a possible progenitor system was found in the \textit{Gaia} ($G=16.587\pm0.004$~mag, $G_{\rm BP}=17.46\pm0.02$~mag, and $G_{\rm RP}=15.40\pm0.01$~mag), 2MASS ($J=13.81\pm0.04$~mag, $H=13.31\pm0.05$~mag, and $K=13.04\pm0.04$~mag), and WISE ($w1=12.58\pm0.02$~mag, $w2=12.52\pm0.02$~mag, $w3=11.0\pm0.3$~mag, and $w4=8.5\pm0.5$~mag) catalogs \citep{2017ATel10774....1P}. According to \textit{Gaia} DR2 \citep{2018A&A...616A...2L}, the distance to the nova could be less than 1~kpc. 
One month after the optical discovery, a weak but significant GeV $\gamma$-ray emission signal was detected at the nova position by \textit{Fermi}-LAT \citep{2017ATel10977....1L}. If the \textit{Gaia} distance is accurate, \src\ will be the closest and also the least luminous $\gamma$-ray nova known to date.

In this paper, we present a multi-wavelength analysis of \src\ based on the observations taken by \textit{Fermi}-LAT, \textit{Swift}/XRT, and the Asiago Novae \& Symbiotic stars (ANS) collaboration. We also discuss whether \src\ can be a sub-luminous $\gamma$-ray nova and, if so, a possible scenario that explains its anomalous features relative to other \textit{Fermi}-detected events.

\section{\textit{Fermi}-LAT Observations}
\label{sec:fermi}
We downloaded the LAT data (Pass 8 with \texttt{P8R3\_SOURCE\_V2} instrument response functions) taken between 8 July 2017 (MJD~57942, the last day of ASAS-SN non-detection) and 4 December 2017 (MJD~58091, day 73 since the first optical detection on MJD~58019) from the \textit{Fermi Science Support Center} (FSSC). 
The analysis tools and the auxiliary database, \texttt{fermitools} (version 1.0.5) and \texttt{fermitools-data} (version 0.17), were used for all of the data reduction and analysis processes. 

The LAT data was selected and reduced based on the analysis threads in FSSC. We chose a region of interest (ROI) of $14\degr\times14\degr$ centred on the nova position. All the events collected within the ROI were further selected so that all events belong to the SOURCE class (i.e., \texttt{evclass=128}) with the event type FRONT or BACK (i.e., \texttt{evtype=3}). 
As Earth's limb is a bright source of $\gamma$-rays to contaminate the LAT observations, all events with zenith angles larger than $90\degr$ are therefore excluded. 
In addition, all the data must be collected during the good time intervals (GTIs) defined by the \texttt{gtmktime} criterion, \texttt{(DATA\_QUAL$>$0)\&\&(LAT\_CONFIG==1)}. 

We used the standard binned likelihood analysis to extract the $\gamma$-ray spectral properties of \src. A spatially-resolved $\gamma$-ray emission model for the ROI was constructed based on the LAT 8-year Source Catalog (4FGL; \citealt{2019arXiv190210045T}). All 4FGL sources located within $20\degr$ from \src\ are included in the emission model. 
All the field sources are at least $1.4\degr$ away from the nova. 
Most of them are much fainter than \src\ according to the flux reported in \cite{2017ATel10977....1L}. 
Therefore, we fixed all the spectral parameters, except for the normalization parameters for the Vela Pulsar (4FGL J0835.3-4510), Vela X (4FGL J0833.1-4511e), and Vela Junior (4FGL J0851.9-4620e), the only three $\gamma$-ray sources that are brighter than \src\ within $5\degr$, for simplicity. 
Besides 4FGL point/extended sources, we employed the background emission models, \texttt{gll\_iem\_v07.fits} and \texttt{iso\_P8R3\_SOURCE\_V2\_v1.txt}, to account for the Galactic diffuse emission and the extragalactic isotropic diffuse emission, respectively. The normalization of the Galactic diffuse component is allowed to vary during fitting, while that of the isotropic diffuse background is fixed at 1, otherwise it will go down to 0.67 (i.e., only 67\% of the default contribution).
We note that fixing the extragalactic isotropic diffuse component do not affect the fitting result significantly--all best-fit parameters of \src\ are well consistent with each other within their 1-$\sigma$ uncertainties.
For our target \src, we assumed a simple power-law model emission model at the ASAS-SN nova position. More complicated models were not considered given the low photon statistics.

\begin{figure*}
\centering
\includegraphics[width=0.9\textwidth]{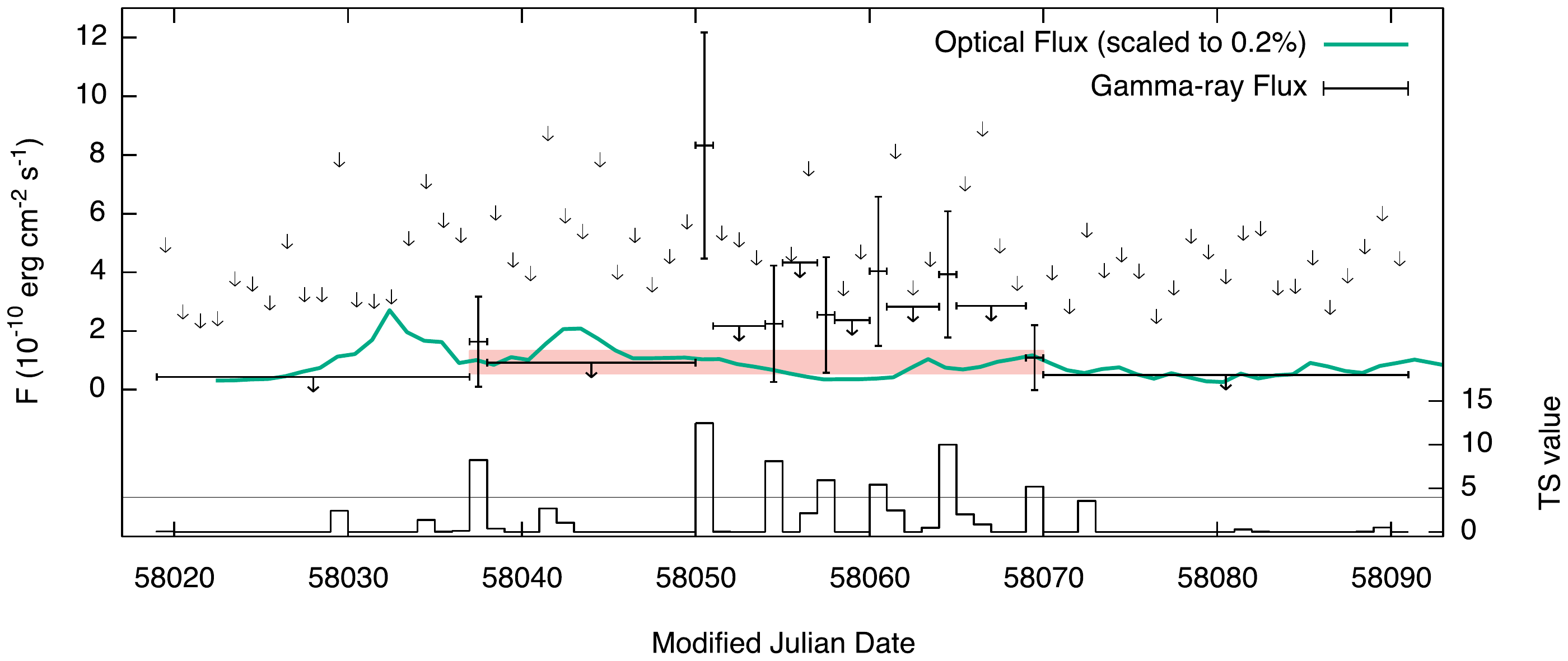}
\caption{The black data bins are the \textit{Fermi}-LAT light curve of \src\ compared with its bolometric light curve (green line). The TS values ($\sqrt{\rm TS}\approx$ the detection significance in $\sigma$) are also shown at the bottom. The black arrows are 95\% upper limits (both daily and stacked are presented) when the nova was undetected (i.e., ${\rm TS}<4$). The red shadow indicates the mean flux level as well as the 1-$\sigma$ uncertainty during the $\gamma$-ray active phase. 
The bolometric flux (green line) was scaled down to 0.2\%, which is the $\gamma$-ray-to-optical luminosity ratio observed in ASASSN-16ma \citep{2017NatAs...1..697L}. 
}
\label{fig:lat_lc}
\end{figure*}

First, we used the likelihood analysis to extract a quick-look LAT light curve with daily bins. The nova was detected in seven bins with the test statistic, ${\rm TS}>4$ (i.e., $\sqrt{\rm TS}\approx$ the detection significance in $\sigma$) with the peak significance at ${\rm TS}>10$ on MJD~58050. Based on the first and last dates of the detections, we define the $\gamma$-ray active phase as MJD~58037--58070 (day 19--52 since the nova discovery). It is worth noting that the $\gamma$-ray flux of the nova was just below the $TS=4$ threshold on MJD~58072 (i.e., $TS=3.6$) when \textit{Swift}-XRT started to observe and detect the nova (further analysis and discussion can be found in \S\ref{sec:xray} and \S\ref{sec:x2g}, respectively). Using the model file described in the previous paragraph, we ran the likelihood analysis with the LAT data collected in this time interval. The nova is significantly detected with ${\rm TS}=33$, equivalent to a detection significance of $>5\,\sigma$. The best-fit photon index and energy flux are $\Gamma_\gamma=1.8\pm0.2$ and $F({\rm 0.1-300GeV})=(9.4\pm3.8)\times10^{-11}$\flux.
We then fixed all spectral parameters of the model file, except the normalizations of the Galactic diffuse emission and \src, and re-extract the LAT light curve (Figure \ref{fig:lat_lc}), which actually looks very similar to the preliminary version. Using the same model and \texttt{gtfindsrc}, the optimized position of the $\gamma$-ray source is (132.581, $-$47.7272; J2000.0) with a 95\% error radius of 5\farcm4. The location is entirely consistent with the nova position with an offset of just 2\farcm5. 

\section{ANS Optical Observations}
\label{sec:ans}
From 26 September 2017 (2 days after the ASASSN discovery) to 8 April 2019 (MJD 58022--58581), the nova was observed nearly daily (when the visibility allows) in Landolt $V$, $B$, and $I$ bands the with ANS Collaboration telescope ID 2100, which is a 40-cm located in San Pedro de Atacama (Chile). Data reduction involves all usual steps of correction for bias, dark, and flat. The transformation for the instantaneous local photometric system to the \cite{2009AJ....137.4186L} equatorial standard system is performed via a highly accurate local photometric sequence, which is pruned and improved with each revisit of the field. With the end of the observing campaign, all observations are re-reduced one final time on the ANS Collaboration central server against the final version of such local photometric sequence, with colour equations ensuing both the highest internal consistency and accurate placing of the data on the Landolt photometric system. All measurements have been performed in aperture mode, with no need to revert to PSF-fitting mode given the low stellar density in the field. Further details on the network of ANS Collaboration telescopes and their operation are provided by \citep{2012BaltA..21...13M} and \citep{2012BaltA..21...22M}.

\begin{figure}
\centering
\vspace{0.2cm}
\includegraphics[width=0.45\textwidth]{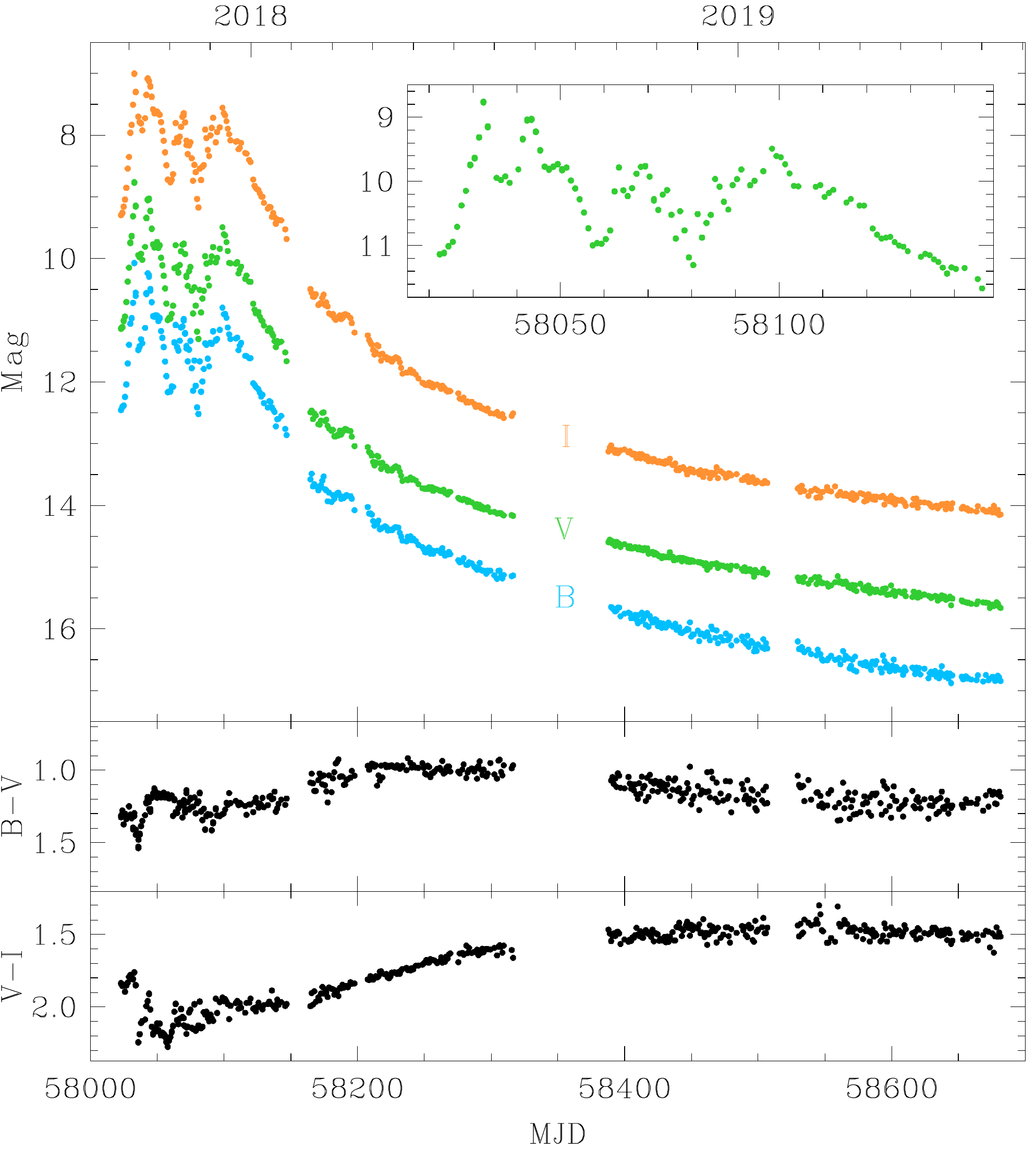}
\caption{ANS Collaboration optical light and colour curves of \src\ (extinction uncorrected). The inset zooms on the early portion of the $V$-band light curve to highlight the presence of large amplitude jitters. 
}
\label{fig:ans}
\end{figure}

Figure \ref{fig:ans} presents the observed $V$, $B$, and $I$-band light curves with the $(B-V)$ and $(V-I)$ evolutions shown in the lower panels.
In the first 100 days of the light curves, \src\ showed at least four jitters of amplitudes of 1--2.5~mag on a time-scale of $\sim1$~month, indicating that it is a J-class nova \citep{2010AJ....140...34S}. 
After 100 days, the nova faded smoothly until the end of the monitoring. Assuming a power-law decline model of $10^{(m_V/-2.5)}\propto(t-t_0)^\alpha$ with $t_0=$ MJD~58020.39 (the ASASSN discovery date), we found that the decline can be well described by a three-segment broken power-law model with the best-fit parameters: $\alpha_1=-3.98\pm0.05$, $\alpha_2=-2.53\pm0.19$, and $\alpha_3=-1.75\pm0.12$, with the two breaks at $(t-t_0)=160\pm4$ and $225\pm19$ days. 

As mentioned in the Introduction (\S\ref{sec:intro}), \src\ could be very nearby. Based on the \textit{Gaia} parallax measurement, the distance to the nova could be $d=560$~pc (see the next section, \S\ref{sec:pre-explo}, for more details), resulting in $M_V=-3.1$~mag as the maximum absolute magnitude of \src\ in the ANS light curve.
In the absolute magnitude distribution of the novae in M31, the absolute magnitude of a nova could be as faint as $M_V\approx-4.8$~mag at maximum (2-$\sigma$ level below the mean value of $M_V\approx-7.2$~mag; \citealt{2017ApJ...834..196S}). However, the obtained absolute magnitude of \src\ is 1.7~mag fainter, likely indicating either a greater distance or the existence of an even brighter nova peak missed prior to the nova discovery.

To allow a direct comparison between the powers radiated in the $\gamma$-ray and optical bands, bolometric correction (BC) was applied on the $V$-band light curve. The BC factors were calculated according to the colour temperatures at different epochs inferred by $T=\frac{7090}{(B-V)_0+0.71}$~K \citep{2007iap..book.....B,2009aste.book.....K,2015arXiv151006262M}. 
We comment that the method may not be ideal as the nova is not a perfect blackbody, but it is still a good approximation for estimating the bolometric fluxes.
The reddening correction was done by assuming a typical $(B-V)_0$ maximum for classical novae, i.e., $(B-V)_0\approx0$ \citep{1987A&AS...70..125V,2016ApJS..223...21H}, which gives $E(B-V)\approx1$ (or $A_V\approx3.1$~mag with $R_V=3.1$). This value indeed agrees well with the one inferred from the \textit{Swift}/UVOT photometry taken in $u$, $uvw1$, $uvm2$, and $uvw2$ bands \citep{2018ATel11649....1P}. 
Figure \ref{fig:lat_lc} shows the resultant bolometric light curve of \src\ (green curve), of which the shape has no big difference from the uncorrected ones.

\section{Pre-explosion optical detection}
\label{sec:pre-explo}
Following the progenitor study by \cite{2017ATel10774....1P}, we searched the \textit{Gaia} \citep{2018A&A...616A...1G}, 2MASS \citep{2006AJ....131.1163S}, and WISE \citep{2010AJ....140.1868W} catalogs at the ASAS-SN position ($08^\mathrm{h}50^\mathrm{m}29\fs576$, $-47\arcdeg 45\arcmin 28\farcs56$) (J2000.0) with a search radius of $1\arcsec$, and found the reported progenitor in all the catalogs. 
The offset is as small as $0\farcs48$ in the \textit{Gaia} map.
We randomly chose an ANS image taken in June 2018 (the nova was $m_V=14$~mag, which is about 10 times brighter than the candidate progenitor) for a better positional constraint. Using 13 bright (i.e., $g<13$~mag) and low proper motion (i.e., slower than $5$~mas~yr$^{-1}$ in both RA and Decl. directions) \textit{Gaia} sources in the field, we construct a \textit{Gaia}-based astrometric solution for the ANS image with an accuracy of $0\farcs37$ (the quadratic sum of the rms uncertainties in RA and Decl. directions). Under this improved solution, the nova is located at $08^\mathrm{h}50^\mathrm{m}29\fs60$, $-47\arcdeg 45\arcmin 28\farcs4$ (J2000.0) and the candidate progenitor is just $0\farcs18$ away from it.
We also noticed a preliminary astrometry for \src\ reported on the \textit{Gaia Photometric Science Alerts} page (identified as Gaia19dff; see \S\ref{sec:intro}), which is entirely consistent with our solution (offset: $0\farcs25$). Because of the unknown accuracy of the preliminary solution, it was not used in the following analysis.

We performed Monte Carlo simulations to test whether these sub-arcsec matches can be easily reproduced by chance. In the simulations, we generated $10^5$ random positions within $2\arcmin$ from the nova, which contains 340 real sources in the \textit{Gaia} catalog. Less than 0.1\% of the simulated stars can find a \textit{Gaia} match within $0\farcs2$, indicating a significance of $>3\sigma$ of the nova-\textit{Gaia} match. 

From the parallax measurement of \textit{Gaia} DR2 (i.e., $\varpi=1.91\pm0.39$~mas, which has been corrected for the global zero point of $-0.029$~mas; \citealt{2018A&A...616A...2L}), the distance to the progenitor source is around $d=560$~pc (the mode value) with a 90\% Bayesian credible interval of 380--1050~pc \citep{2018AJ....156...58B}, indicating that V549 could be the closest $\gamma$-ray emitting nova known to date. 
However, a large astrometric excess noise of $\epsilon_i=3.2$~mas is also presented in the catalog, showing that there is a significant difference between the data and the astrometric solution. Nevertheless, we take $d=560$~pc as a reference and further discussions on the issue will be given in \S\ref{sec:gaia} and \S\ref{sec:ms}.

The progenitor source is not particularly bright with $m_B=17.4$~mag (NOMAD; \citealt{2004AAS...205.4815Z}), equivalent to an absolute magnitude of $M_B=3.2$--5.4~mag for $d=380$--1050~pc, if $E(B-V)=1$ inferred in the previous section is applied.

It is well known that cataclysmic variables (CVs) are the progenitors of classical novae, and thus the \textit{Gaia} progenitor would probably be a CV, if \src\ is nearby.
If the accretion is active, large variability can be seen in CVs.
In the WISE catalog (43 epochs in MJD~55343--45 and 55533--35), \src's progenitor is classified as a stable system with \texttt{Variability flag = 2} and \texttt{1} in the $w1$ and $w2$ bands, respectively (i.e., a source is considered stable if the flag is less than 6).
Besides, no previous outburst of \src\ has been reported in the literature.
The stability might suggest a low accretion rate of the system. 
Moreover, CVs with high accretion rates are bright because of the accretion disk. For example, V392 Per, which also exhibited as a $\gamma$-ray nova in 2018 \citep{2018ATel11590....1L}, has \textit{V}-band absolute magnitudes from $-0.2$ to 2.3~mag during the quiescence \citep{2018RNAAS...2...24D,2020A&A...639L..10M}. 
If \src\ had the same quiescent level, the extinction has to be $A_V=3.8$--$7.4$~mag ($d=560$~pc is assumed), which is higher than the values obtained from the ANS (i.e., $A_V\approx3.1$~mag).
We therefore argue that the accretion rate of \src\ was relatively low in quiescence. In this context, the companion stellar emission could dominate in the pre-nova observations.

\begin{figure*}
\centering
\includegraphics[width=0.49\textwidth]{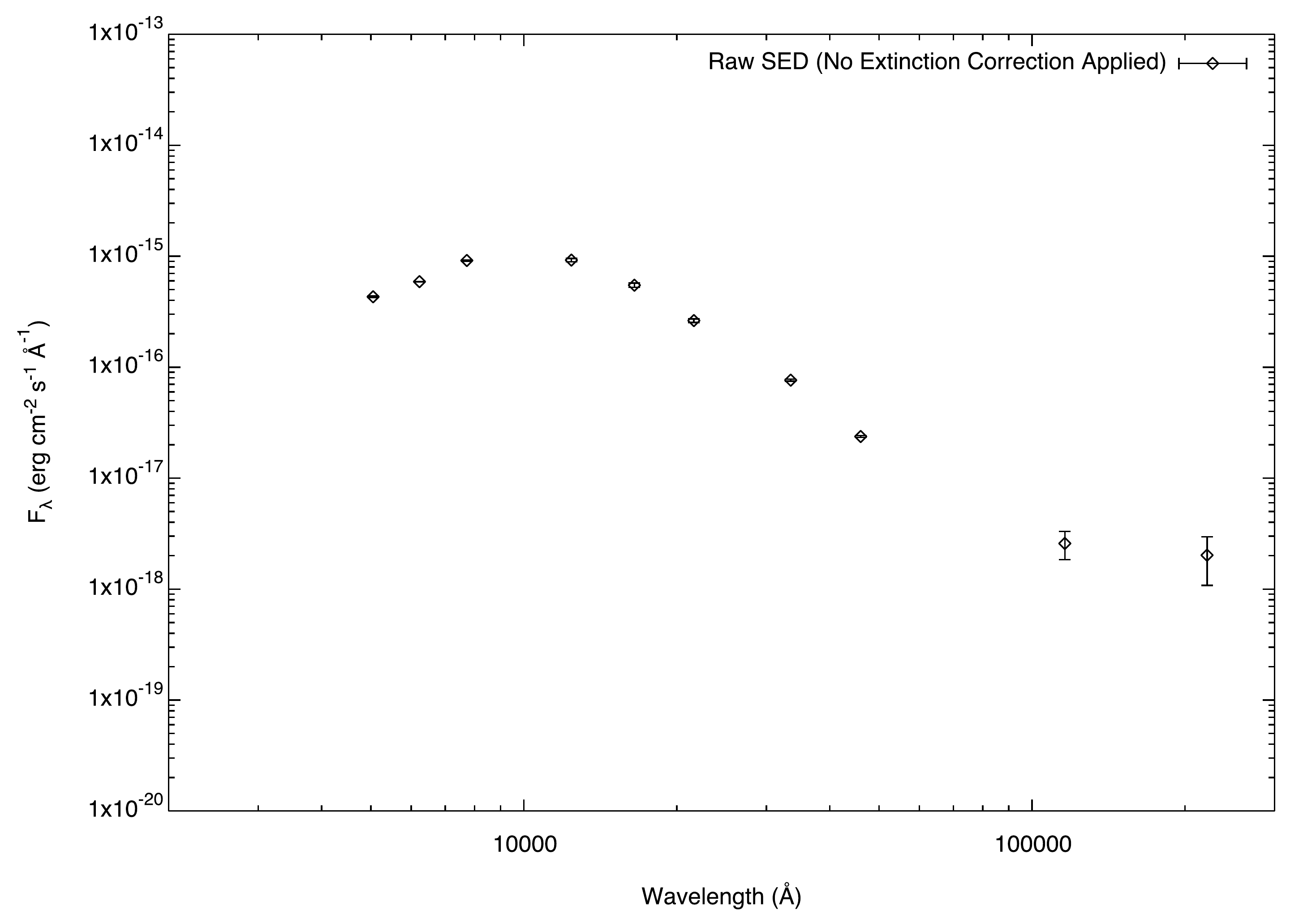}
\includegraphics[width=0.49\textwidth]{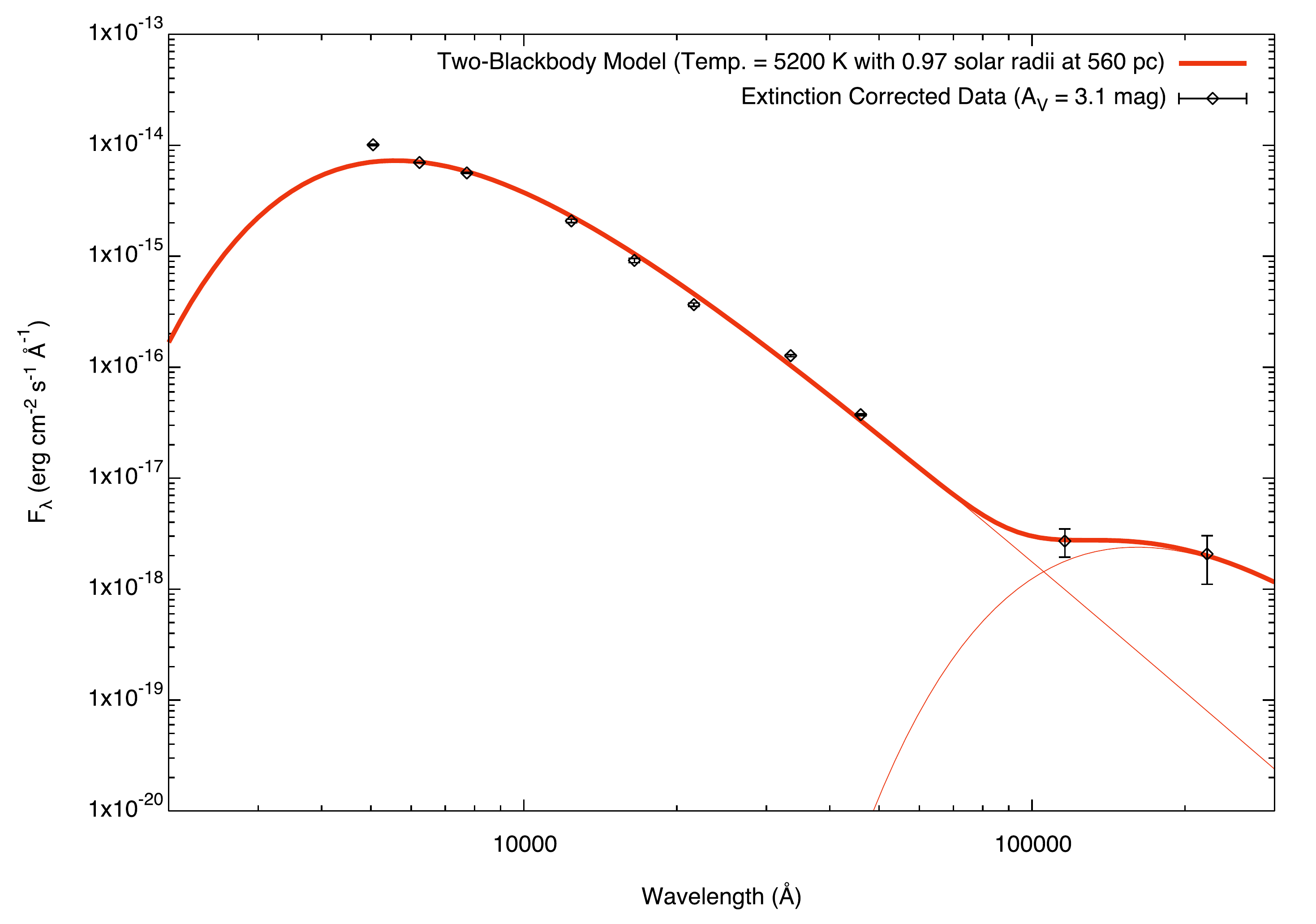}
\includegraphics[width=0.49\textwidth]{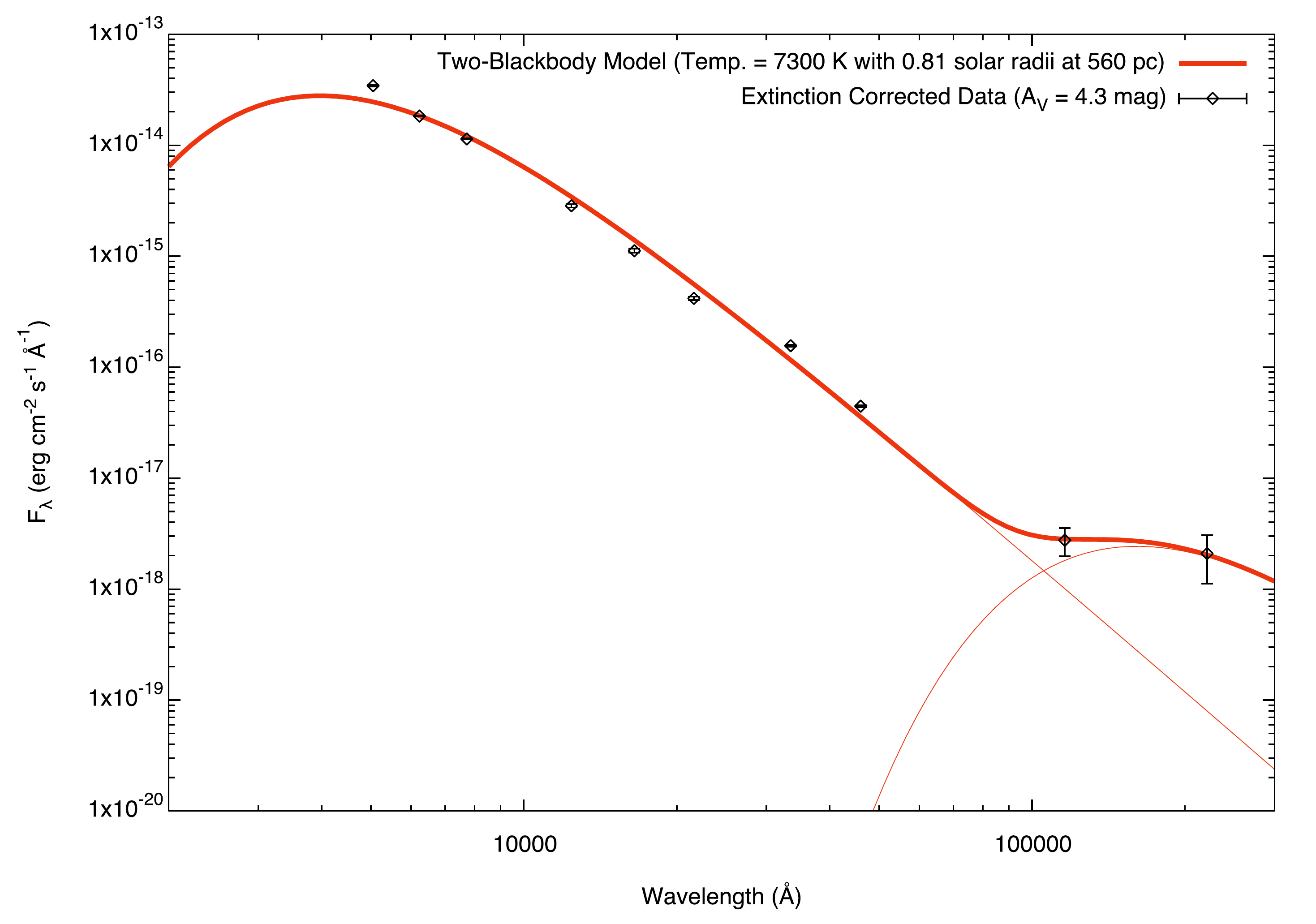}
\includegraphics[width=0.49\textwidth]{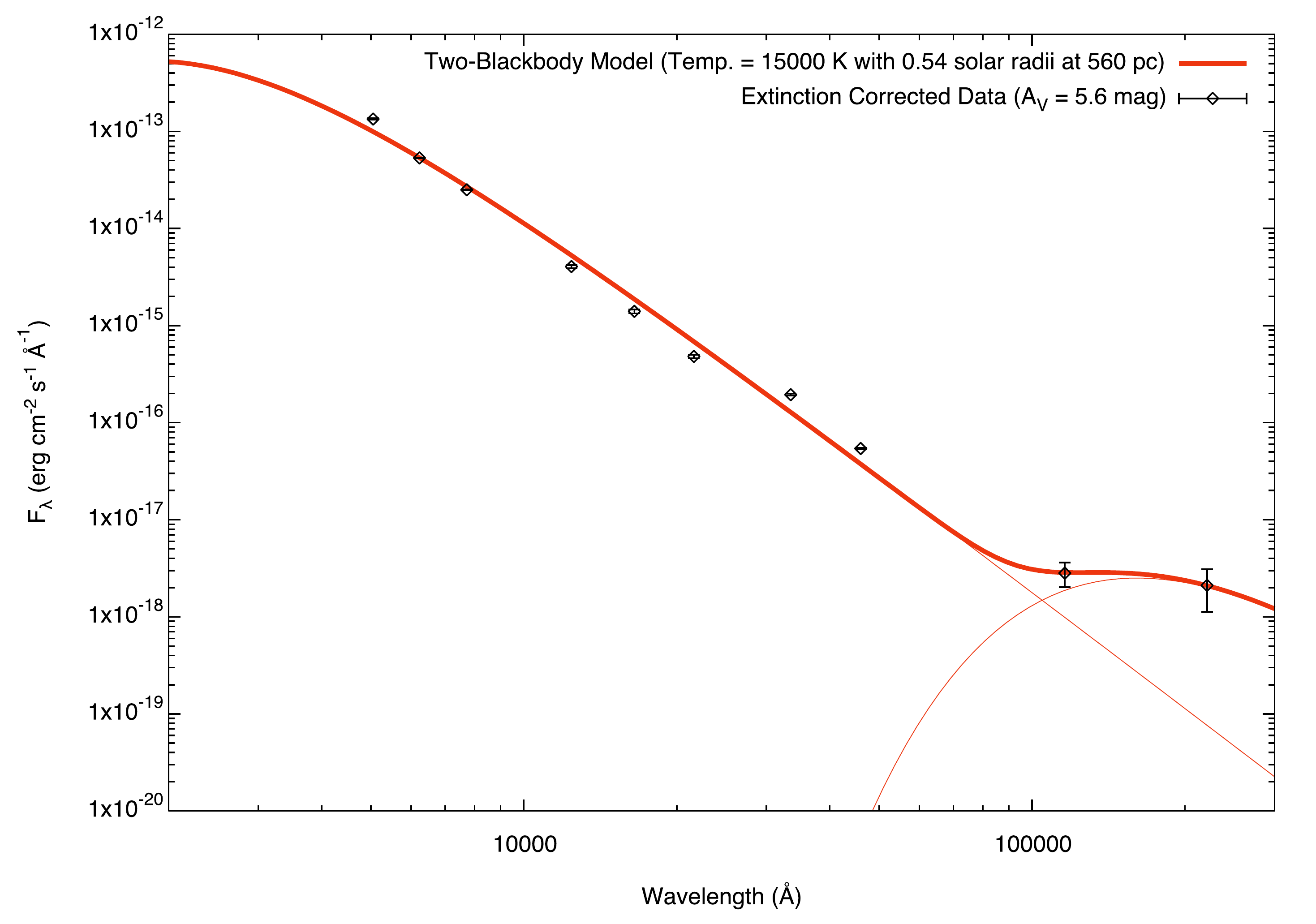}
\caption{The upper-left figure shows the uncorrected SED of the progenitor of \src, while the rest are the extinction corrected SEDs with the best-fit two-blackbody models (red line). The warm blackbody components were all around $\approx200$~K in temperature and $\approx80\,R_\sun$ (or 0.4~au) in radius.
}
\label{fig:sed}
\end{figure*}

By assuming the progenitor source as a blackbody, we measured the surface temperature as well as the size of the companion using the spectral energy distribution (SED) comprised of the data from \textit{Gaia}, 2MASS, and WISE (Figure \ref{fig:sed}). 
Three extinction curves, \cite{1989ApJ...345..245C,2013MNRAS.430.2188Y,1999PASP..111...63F}, were used to deredden the SED in the optical/near-infrared (IR), mid-IR, and far-IR bands, respectively. 
Besides the extinction value $A_V=3.1$~mag deduced in \S\ref{sec:pre-explo}, we also adopted $A_V=3.1$ (duplicated), 4.3, and 5.6~mag inferred from the lower-bound, best-fit, and upper-bound hydrogen column densities (i.e., \nh\ $=9.5^{+2.9}_{-2.7}\times10^{21}$\cm; converted with $A_V = \frac{N_\mathrm{H}}{2.21\times10^{21}\rm{cm}^{-2}}$~mag; \citealt{2009MNRAS.400.2050G}) obtained from the X-ray spectral analysis during the supersoft X-ray phase (see the next section, \S\ref{sec:xray}, for a more complete analysis). We note that the \nh\ value was measured in a very late phase of the eruption (days 253--389), during which the contribution from the ejecta should be limited. In the WISE $w3$ ($12\,\mu$m) and $w4$ ($22\,\mu$m) bands, there is a clear IR emission excess (Figure \ref{fig:sed}), which is indicative of warm dust around the star, probably a circumbinary disk around the CV \citep{2001ApJ...548..900S,2001ApJ...561..329T}. An additional blackbody was therefore included. In all of the fits, the IR excess can be well modelled by a warm blackbody of $T_{\rm {dust}}\approx200$~K and $R_{\rm {dust}}\approx80\,R_\sun$ (or 0.4~au; $d=560$~pc assumed). No uncertainty is given as there is no degree of freedom for the IR excess model. 

The extinction corrected SEDs can be well described by blackbody radiation. The best-fit temperatures are $5200\pm200$~K (for $A_V=3.1$~mag), $7300\pm500$~K (for $A_V=4.3$~mag), and $15000\pm3000$~K (for $A_V=5.6$~mag) with radii of $0.97\pm0.08\,R_\sun$, $0.81\pm0.09\,R_\sun$, and $0.54\pm0.11\,R_\sun$, respectively, if $d=560$~pc. 
The SEDs are in general slightly brighter than the models at the blue end, possibly indicating an accretion component.
Among the fits, the $A_V=3.1$~mag set is consistent with a Sun-like star, and the corresponding extinction also matches well with the reddening inferred from the ANS light curves, i.e., $E(B-V)\approx1$ (see \S\ref{sec:ans}). Nevertheless, we note that the SED models considered here do not take the accretion, however weak, into account, and therefore the stellar types and luminosity classes inferred could vary if an improved model is used.

\begin{figure*}
\centering
\includegraphics[width=0.49\textwidth]{sed_noextinction_v3.pdf}
\includegraphics[width=0.49\textwidth]{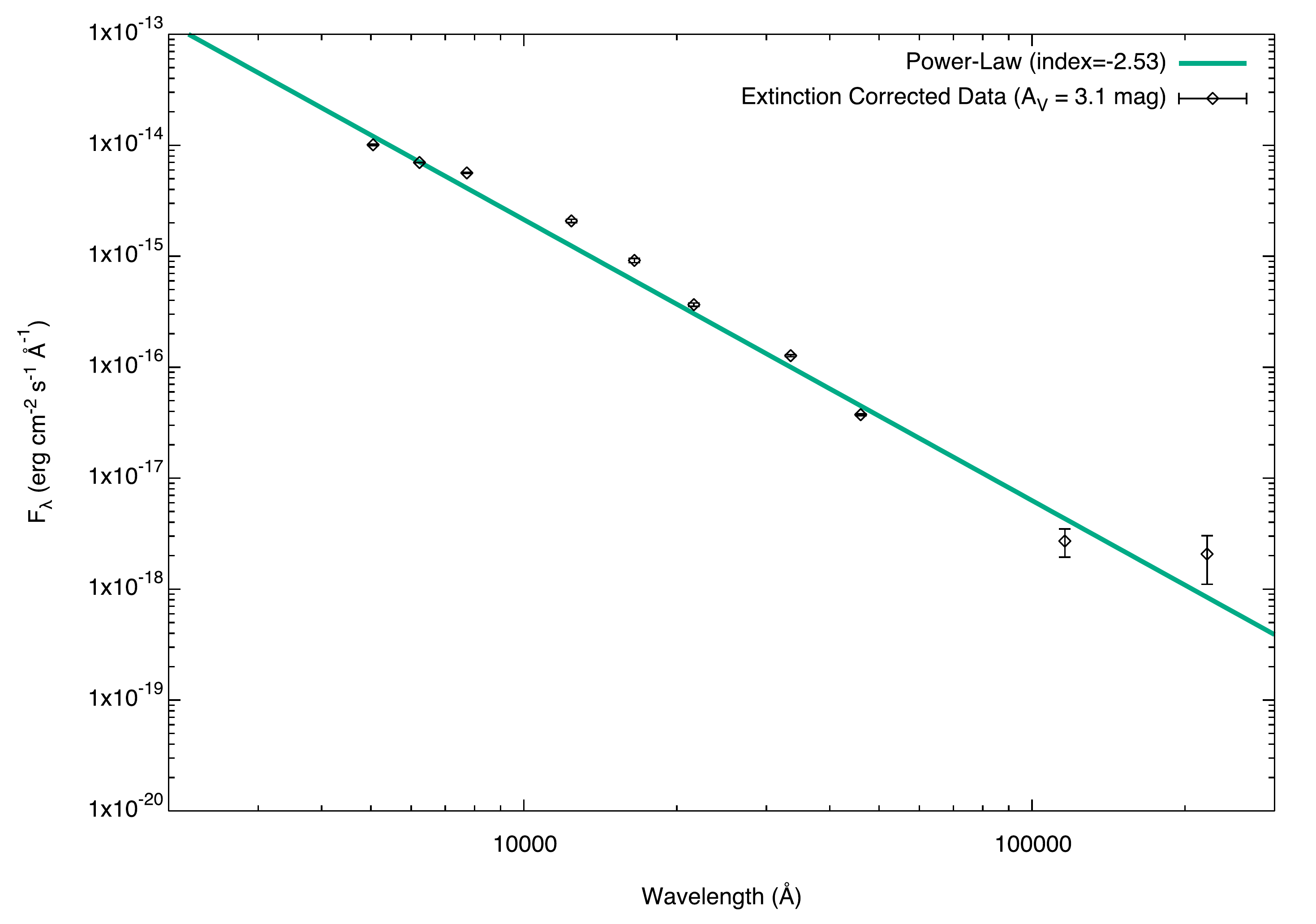}
\includegraphics[width=0.49\textwidth]{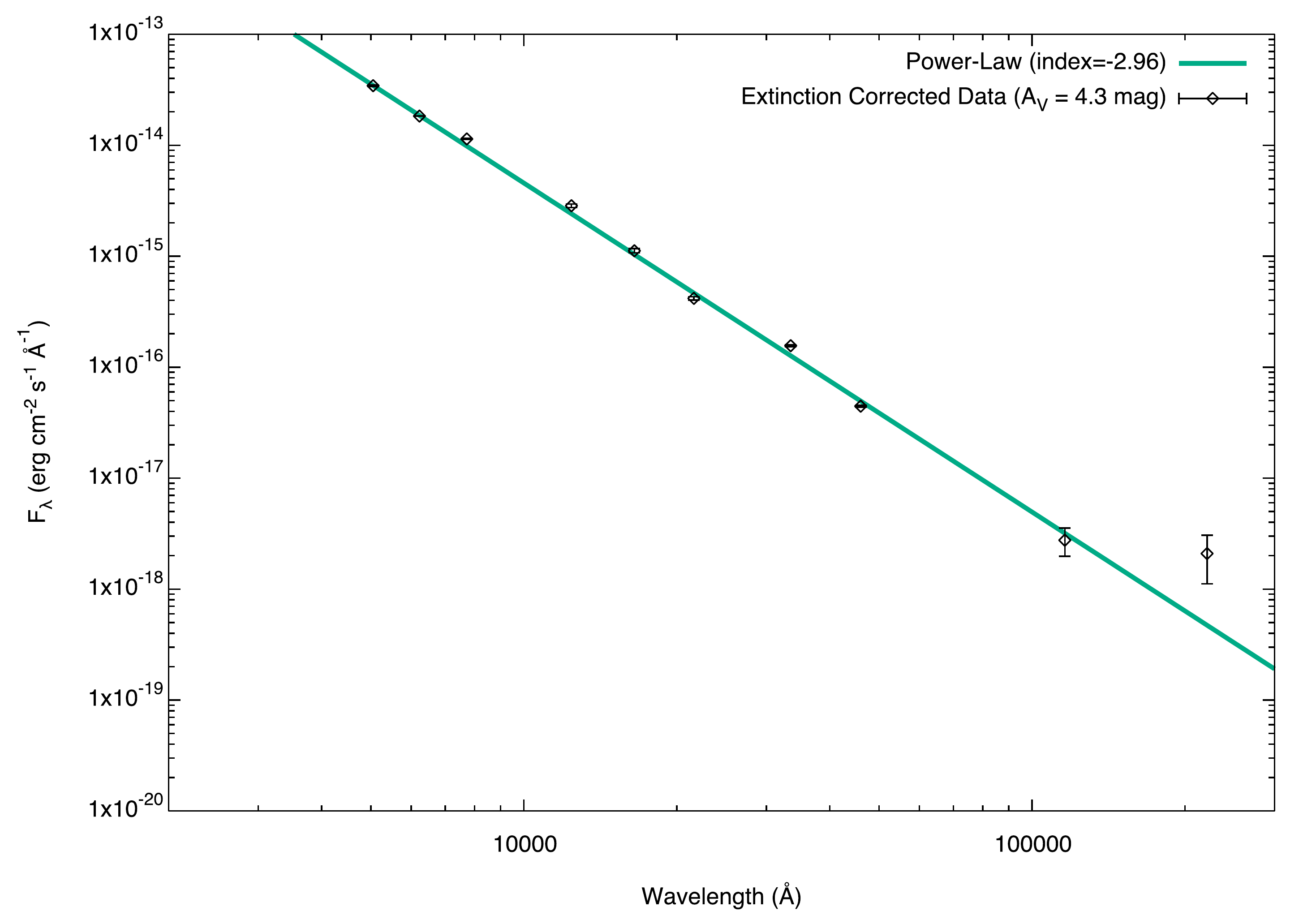}
\includegraphics[width=0.49\textwidth]{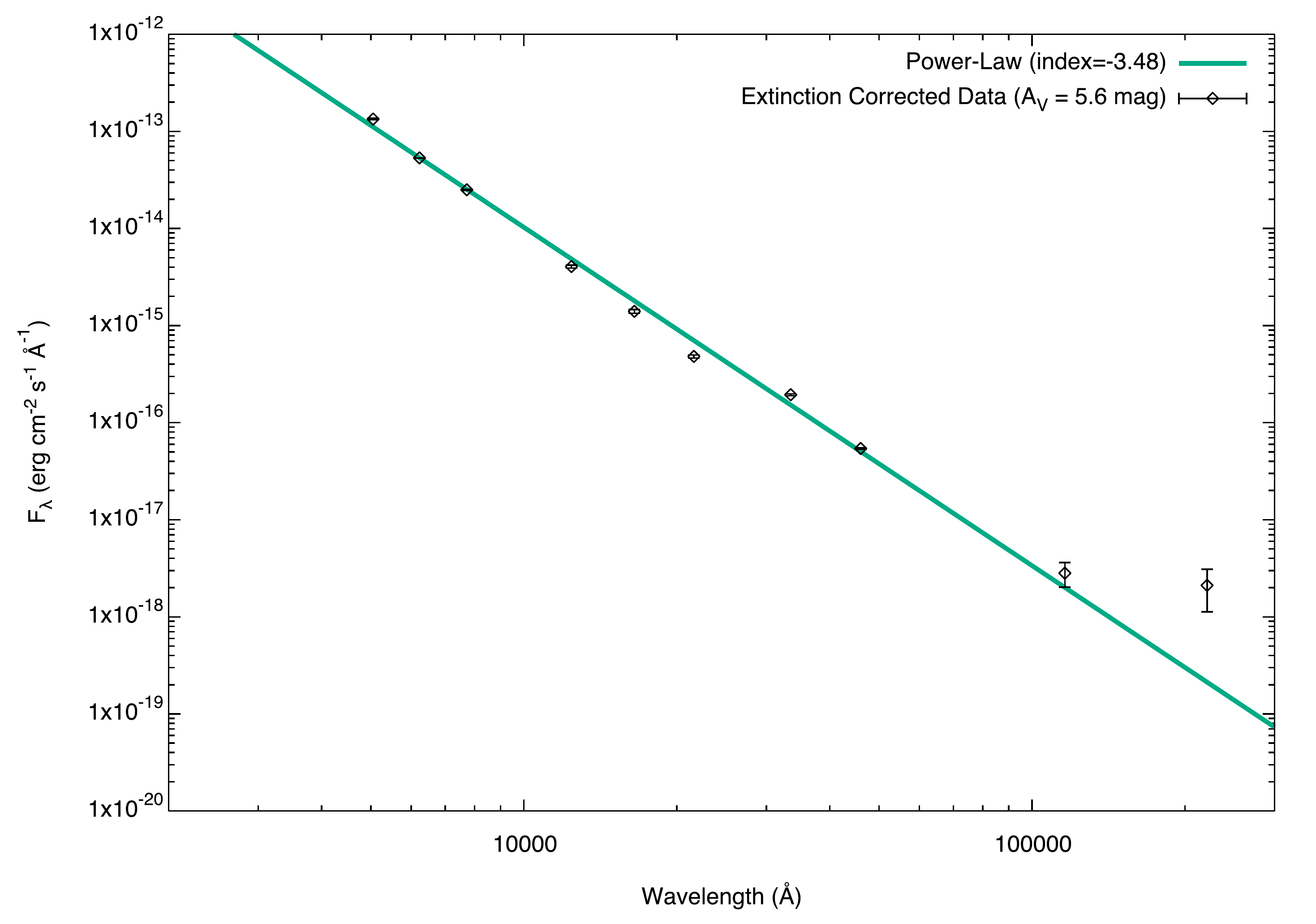}
\caption{Same as Figure \ref{fig:sed}, but with best-fit power-law models.
}
\label{fig:sed2}
\end{figure*}

We also fitted the SEDs with a power-law model to explore other possibilities, e.g., an accretion-dominated system with hot plasma. With the power law, the IR excess is marginally seen only in the $w4$ band. We therefore skipped the warm blackbody component for simplicity. The best-fit spectral indices are $-2.53\pm0.08$ (for $A_V=3.1$~mag), $-2.96\pm0.05$ (for $A_V=4.3$~mag), and $-3.48\pm0.06$ (for $A_V=5.6$~mag).
In general, the two SEDs with higher extinctions give very good fits to the power-law model. While the fit for $A_V=3.1$~mag is reasonably good, the SED is slightly curved in the log-log plot, and does not fully follow a power law (Figure \ref{fig:sed2}). Given that $A_V=3.1$~mag is likely the correct extinction as it is consistent with both the ANS and X-ray analyses, we slightly favour the blackbody fit that provides a better description for the SED, if $A_V=3.1$~mag.

\section{\textit{Neil Gehrels Swift} XRT Observations}
\label{sec:xray}
\textit{Swift}-XRT has observed \src\ 33 times during the 2017 outburst (from MJD~58072 to 58427), and the nova was clearly detected in the XRT's energy domain \citep{2018ATel11649....1P}. The X-ray emission was first detected on day 54 (MJD~58072) since the ASAS-SN discovery with a plasma temperature of 1--2~keV based on the APEC model. It is important to note that the nova was also marginally detected with ${\rm TS}=3.6$ by \textit{Fermi}-LAT on the same day. 
Since day 236 (MJD~58254), supersoft X-ray emission with a blackbody temperature of $kT\approx30$~eV was also seen until the last XRT observation taken on day 409 or MJD~58427 \citep{2018ATel11649....1P,2019arXiv190802004P}. 

We downloaded the \textit{Swift}-XRT observations and re-analysed parts of the dataset to aim for a better understanding of the system. In particular, we wish to make constraints on the nova distance based on the supersoft thermal X-ray emission and measure the flux ratio between the X-rays and $\gamma$-rays on day 54 (around MJD~58072). 

\begin{figure*}
\centering
\includegraphics[width=0.85\textwidth]{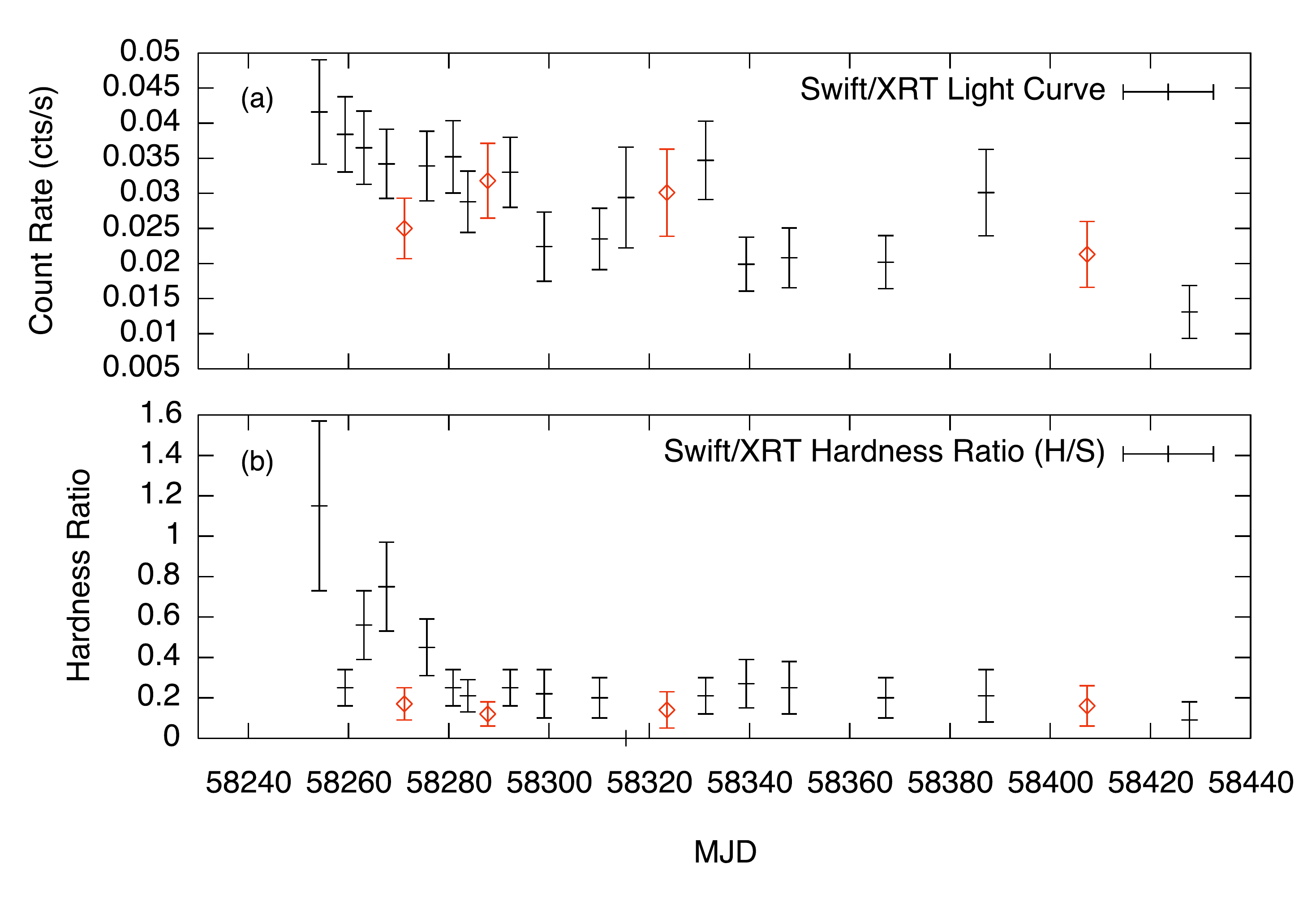}
\caption{The X-ray light curve and the hardness ratio evolution of \src\ during the supersoft X-ray phase. The red data points represent the observations we used for the supersoft X-ray analysis (see text for details). 
}
\label{fig:xrt_lc}
\end{figure*}

\texttt{HEAsoft} (version 6.24) was used to reduce and analyse the XRT data with the HEASARC's calibration database (CALDB; version 20180710). 
The light curve and spectra were extracted using the task \texttt{xrtgrblc} that applies count-rate dependent regions to optimize the signal-to-noise ratio (S/N) of the products. The obtained light curve (0.3--10~keV) is consistent that shown in \cite{2019arXiv190802004P}. We also extract a hardness ratio curve (i.e., hard-to-soft X-ray count ratio with 1.5~keV as the energy cut) for the data selection that will be described in the next paragraph. Figure \ref{fig:xrt_lc} shows the light and hardness ratio curves of \src\ in the supersoft X-ray phase. 

We first focused on the spectral analysis of the supersoft emission. As the photon statistics per individual observation is not great, we performed stacked analysis to estimate the blackbody temperature as well as its size. 
While the supersoft X-ray emission is fairly stable throughout the observed period, only those observations with similar X-ray hardnesses and count rates were used to ensure that all stacked data were taken in very similar spectral states. The selection criteria are ``\texttt{0.2 $>$ hardness $>$ 0.1}'' and ``\texttt{0.04~cts/s $>$ count rate $>$ 0.02~cts/s}''. We note that the boundaries of the criterion are decided rather arbitrarily and selection bias could exist, but the result should still be sufficient to check whether the supersoft X-ray emitting region has a sensible size at the \textit{Gaia} distance. Four observations (days 253, 269, 305, and 389; or MJD~58271, 58287, 58323, and 58407) were finally selected and the total exposure time is about 6~ksec. 

\texttt{XSPEC} was used for the spectral fitting. Since the total number of counts in the four observations (including background) is only 117, \textit{C-Statistic} is used as the fit statistic and we binned the spectrum to at least one count per bin as suggested by the \textit{Swift} team. The stacked spectrum can be well fitted with \texttt{phabs*(apec+bbodyrad)} that was also used in the analysis of \cite{2018ATel11649....1P}. 

Strong absorption was found with \nh\ $=9.5^{+2.9}_{-2.7}\times10^{21}$\cm. For the APEC component, the best-fit temperature is $1.3^{+0.4}_{-0.3}$~keV with an unabsorbed flux of $8.4\times10^{-13}$\flux\ (0.3--10~keV). The supersoft X-ray component has a blackbody temperature of $kT_{\rm bb}\approx 37$~eV and an unabsorbed flux of $5.2\times10^{-9}$\flux\
in 0.3--1~keV, equivalent to an emission size of $R_{\rm bb}=5^{+8}_{-3}\times10^8\,(d/560\,{\rm pc})$\,cm in radius. 
The inferred blackbody temperature is consistent with $kT_{\rm bb}=28^{+57}_{-19}$~keV presented in \cite{2018ATel11649....1P}.
To elaborate the uncertainty estimation, 90\% confidence limits were computed using the \texttt{XSPEC} command \texttt{error} with the blackbody temperature fixed at the best-fit value. If the $T_{\rm bb}$ parameter is not frozen, the lower bound of $R_{\rm bb}$ will keep hitting the zero ground in the calculation. If we let the code continue, it gives $kT_{\rm bb}=37^{+13}_{-8}$~keV and $R_{\rm bb}<24\times10^8\,(d/560\,{\rm pc})$\,cm (90\% confidence).

For the first XRT spectrum taken on MJD~58072 (when the $\gamma$-ray emission was marginally detected on day 54), a single-component APEC model is enough to give a good fit with \nh\ $=1.8^{+0.6}_{-0.5}\times10^{22}$\cm\ and $kT=2.3^{+1.0}_{-0.5}$~keV. 
The fitting result is consistent with that of \cite{2018ATel11649....1P}.
Based on these parameters, the unabsorbed flux is inferred to be $8.0^{+4.7}_{-2.2}\times10^{-12}$\flux\ (0.3--10~keV). Compared with the marginal \textit{Fermi}-LAT detection of $F({\rm 0.1-300GeV})=(1.6\pm1.5)\times10^{-10}$\flux, the X-ray-to-$\gamma$-ray flux ratio is about 5\%. The 1-$\sigma$ confidence interval is roughly 2--77\%, which is estimated by the possible range allowed by the 68\% uncertainties of the corresponding X-ray and $\gamma$-ray fluxes. 

\section{Discussion and conclusion}

\subsection{Thermal X-ray emission from shocks}
\label{sec:x2g}
While supersoft X-ray emission of novae originates from the white dwarf atmosphere, hard X-rays ($\gtrsim1$~keV) are generally believed to come from shocks (see, e.g., \citealt{2001A&A...373..542O}). These shocks can be either external shocks formed when the nova ejecta runs into the circumstellar medium, or internal shocks formed when the nova outflows of different speeds collide. For $\gamma$-ray emitting novae, \cite{2014MNRAS.442..713M} theoretically showed that thermal hard X-rays can be produced in the internal shocks that radiate the observed $\gamma$-ray emission. In this sense, concurrent X-ray and $\gamma$-ray observations of a nova could put crucial constraints on the radiative efficiency and also the particle acceleration efficiency of the internal shocks. 

In \src, the quasi-simultaneous \textit{Swift} and \textit{Fermi}-LAT flux measurements taken on MJD~58072 result in a flux ratio of 2--77\%, unambigously showing that the $\gamma$-ray luminosity is higher. This, despite the difference in the X-ray energy ranges, is in agreement with the result of the recent \textit{Fermi}-LAT and \textit{NuSTAR} concurrent observations of the two novae V5855~Sgr and V906~Car, which yield $\sim$0.1--1\% \citep{2019ApJ...872...86N} and 2\% \citep{2020MNRAS.497.2569S,2020NatAs.tmp...79A}, respectively. However, in the nova shock model of \cite{2014MNRAS.442..713M} with standard parameters assumed, one would expect to observe much brighter thermal X-ray emission---10 to 1000 times more luminous than the $\gamma$-ray flux---and this contradiction was also found in V5855~Sgr and V906~Car. 

\cite{2019ApJ...872...86N} and \cite{2020NatAs.tmp...79A} argued that the low X-ray luminosity indicates an X-ray suppression in the shocks, which could be due to thin-shell instabilities at the shock front. In the standard one-dimensional solution, the post-shock temperature can be written as $T_{\rm sh}\approx1.4\times10^7(v_{\rm sh}/1000\,{\rm km\,s^{-1}})^2$\,K for nova shocks ($v_{\rm sh}$ is the shock velocity here; \citealt{2018MNRAS.479..687S}), which is around $T_{\rm sh}\approx10$~keV for $v_{\rm sh}={\rm a~few\times10^3}\,{\rm km\,s^{-1}}$. However, in multi-dimensional cases, thin-shell instabilities would generate a corrugated shock structure, which can significantly suppress the thermal X-rays radiated from the shocks \citep{2018MNRAS.479..687S}. 
This suppression mechanism could also explain the low X-ray-to-$\gamma$-ray flux ratio of \src. 
In the quasi-simultaneous \textit{Swift} observation, \src\ showed a plasma temperature of $2.3^{+1.0}_{-0.5}$~keV that is a few times lower than the value expected for a planar shock (i.e., $\approx10$~keV), supporting the scenario. Considering also the fact that all the three \textit{Fermi}-detected novae that were observed (quasi-)simultaneously in X-rays and $\gamma$-rays have similar low X-ray fluxes, it is likely that the X-ray suppression is very common in the nova shocks that radiate observable $\gamma$-rays. 

Some of the above arguments are in part based on a marginal \textit{Fermi}-LAT detection of \src\ (i.e., $\lessapprox2\,\sigma$). Because of the low significance, the $\gamma$-ray flux inferred by the likelihood analysis could be inaccurate. If the detection is not real, the actual $\gamma$-ray flux would be much lower and the X-ray luminosity could be more consistent with the theoretical prediction with no suppression. Unfortunately, the quasi-simultaneous \textit{Swift} data was the first X-ray data taken for \src, and therefore no further X-ray constraint can be placed on the system when it was brighter in $\gamma$-rays. 

Another interesting X-ray property of \src\ is the low \nh\ value observed during the $\gamma$-ray active phase (i.e., \nh~$\sim10^{22}$\cm) compared to that of V5855~Sgr and V906~Car, which are up to \nh~$\sim10^{24}$\cm. In \cite{2014MNRAS.442..713M}, internal shocks are formed when a fast nova outflow interacts with a dense external shell (DES). In the case of classical novae (e.g., ASASSN-16ma \citealt{2017NatAs...1..697L}), the DES usually refers to a slow and dense outflow launched during the earlier phase in the eruption \citep{2015MNRAS.450.2739M}. We argue that the much lower \nh\ of \src\ implies a very different physical origin of the DES in the system (i.e., not a slow ejecta). It could be related to the ambient dust that causes the IR excess found in the SED of the progenitor (Figure \ref{fig:sed}; see \S\ref{sec:pre-explo}), which will be discussed in \S\ref{sec:ms}. 

\subsection{Uncorrelated optical and $\gamma$-ray light curves}

In ASASSN-16ma, \cite{2017NatAs...1..697L} found a strong correlation between the optical and $\gamma$-ray light curves, which shows the dominance of the shock emission in the optical band (see also the analysis and discussion in \citealt{2017MNRAS.469.4341M}). The recent \textit{Fermi}-LAT and \textit{BRITE} \citep{2016PASP..128l5001P} observations of V906~Car (also known as ASASSN-18fv) confirmed such a correlation, and hence, showed that the shock optical emission is common in nova eruptions \citep{2020NatAs.tmp...79A}. Using the \textit{Fermi}-LAT observation and the photometric data from ANS, we have made a direct comparison between the $\gamma$-ray and bolometric corrected optical light curves (Figure \ref{fig:lat_lc}). Unlike ASASSN-16ma and V906~Car, the emission of \src\ in the energy bands are not correlated at all. The deviation is particularly clear in the first month since the explosion (i.e., about MJD~58020--58030 in Figure \ref{fig:lat_lc}), where two bright optical flares were observed but no significant $\gamma$-ray emission is detected during the periods. Although there was a daily \textit{Fermi}-LAT detection on MJD~58037, it sits right at the local minimum between the two optical flares and does not look like to be associated with the optical flux. 

The uncorrelated optical and $\gamma$-ray light curves of \src\ indicates that the majority of the optical emission is not shock-powered. 
We speculate that the shocks still emitted optical emission, but the flux was just too weak to take over in the band.

When the nova was optically fainter in MJD~58037 and 58050--58070, the average $\gamma$-ray emission was in the order of $\sim0.1$\% of the optical flux, which is comparable to the AAVSO optical and \textit{Fermi}-LAT $\gamma$-ray observations of ASASSN-16ma that finds $L_{\rm \gamma}/L_{\rm opt}\approx0.2\%$ \citep{2017NatAs...1..697L}. The shock emission could contribute significantly to the observed ANS optical light curve in these time intervals, during which a correlation between the optical and $\gamma$-ray light curves could be possibly observed. 
But still, no obvious correlation can be seen in the period, either because the S/N of the LAT data is too low, or the actual shock emission fraction was much higher than $\sim0.1$\% in $\gamma$-rays (i.e., $L_{\rm \gamma}/L_{\rm opt}\gg0.1\%$).

\subsection{The \textit{Gaia} distance measurement}
\label{sec:gaia}

The pre-nova \textit{Gaia} parallax measurement of \src\ indicates that the nova could be as nearby as $d=380$--1050~pc (90\% Bayesian credible interval), with a posterior mode of $d=560$~pc, to us \citep{2018AJ....156...58B}. 
Adopting $d=560$~pc, the \textit{Gaia} counterpart could be a low-mass star of G type 
(we caution that the result is highly sensitive to the extinction as well as the SED model adopted).
At the same distance, the supersoft X-ray emission observed by \textit{Swift}-XRT would have a blackbody size of $R_{\rm bb}\sim5\times10^{8}$~cm in radius, which is fully consistent with that of some well-known systems, e.g., RS Oph with $R_{\rm bb,RS}\sim7\times10^{8}$~cm observed in the 2006 outburst (\citealt{2011ApJ...727..124O}; note also that, if $R_{\rm bb}<24\times10^8\,(d/560\,{\rm pc})$~cm presented in \S\ref{sec:xray} is considered, a greater distance to \src\ could also lead to a consistent result). All these observational facts are suggestive of a nearby classical nova residing within 1~kpc of the Sun.

On the other hand, the maximum magnitude of \src\ in the ANS light curve is too faint for a regular nova at $d=560$~pc. In addition, given the presumed proximity, the foreground Galactic extinction should be low (e.g., $E(B-V)=0.05$, estimated by the Stilism 3D dust map; \citealt{2014A&A...561A..91L,2017A&A...606A..65C}), contradicting the reddening of $E(B-V)\approx1$ derived by the previous analyses.

The reliability of the $Gaia$ distance measurement is certainly a major factor on the issue. As mentioned in \S\ref{sec:pre-explo}, there is a large excess noise of $\epsilon_i=3.2$~mas in the standard five-parameter astrometric solution of \src\ (i.e., positions, proper motions, and parallax; see the sections 3.6 and 5.1.2 of \citealt{2012A&A...538A..78L} for details), indicating a large discrepancy between the data and the solution.

In \textit{Gaia} DR2, all the cataloged sources were assumed to be single stars, and therefore astrometric excess noises can be found in unresolved binaries due to the orbital motions \citep{2018A&A...616A...2L}.
Yet, for $d=560$~pc, $\epsilon_i=3.2$~mas is equivalent to 380$\,R_\sun$, which is too large for the orbit of a CV (i.e., in the order of $R_\sun$). Alternatively, the progenitor could be a symbiotic binary located at a much farther away distance, and the obtained \textit{Gaia} parallax is merely the result of the orbital motion of the giant companion. 
In this scenario, the tensions regarding the faintness of the nova and the large extinction observed would be immediately relieved either.
However, if the distance to the nova is, e.g., $d>5.6$~kpc (to make the radius of the giant as $>10\,R_\sun$), $\epsilon_i=3.2$~mas and $\varpi=1.91$~mas will refer to $>3800\,R_\sun$ and $>2300\,R_\sun$, which are also too large for a typical symbiotic binary (the projected size is smaller than a few hundred $R_\sun$ with orbital periods of 200--1000 days; see, e.g., \citealt{2003ASPC..303....9M}).
In some extreme symbiotic systems, the orbital sizes can reach 10,000$\,R_\sun$ (e.g., V407~Cyg; \citealt{2013ApJ...770...28H,2020A&A...638A.130G}), but their orbital periods are too long (i.e., $\sim100$~years) to mimic the parallax effect that is on a yearly time-scale.

Another possibility is that the astrometric excess noise is not due to the binary nature of the system, but instead, because of the extended feature that we have observed in the WISE data. Assuming $d=560$~pc, the feature would have a radius of $\sim80\,R_\sun$, which is more comparable to the excess noise. It is possible that the scattered light from the warm dust affect the point spread function (PSF) of \src's progenitor to cause some of the excess noise, although the significance of the effect is unclear.

The \textit{Gaia} DR2 parallax for \src\ is noisy, yet some possible scenarios for the origin of the astrometric excess noise are consistent with a nearby distance.
Depending on the distance, \src\ would be a rare sub-luminous $\gamma$-ray nova or a more ordinary symbiotic system. The former case, which needs further investigations on the implications, will be discussed in the next section (with $d=560$~pc as a reference).

\subsection{If \src\ is nearby}
\label{sec:ms}

\begin{table*}
\scriptsize
\centering 
\caption{The $\gamma$-ray luminosities of all the $\gamma$-ray novae with estimated distances in the literature.}
\begin{tabular}{llcccl}
\toprule
Name		&	Distance estimation method	&	Distance	&	$L_\gamma$		&	$L_\gamma/L_{\rm V549}$\footnote{$L_{\rm V549}=4\times10^{33}$\lum, which is the $\gamma$-ray luminosity of \src\ assuming $d=560$~pc.}	&	References \\
		&					&	(kpc)		&	(\lum)			&				&		 \\
\hline
V5668~Sgr	&	NIR analysis on the dust shell 	&	2.0		&	$3.0\times10^{34}$	&	8.5			&	﻿\cite{2016ApJ...826..142C,2016MNRAS.455L.109B} \\
V959~Mon	&	High-resolution radio imaging			&	1.4		&	$5.6\times10^{34}$	&	16			&	﻿\cite{2014Sci...345..554A,2015ApJ...805..136L} \\
V1369~Cen	&	High-resolution spectroscopy	&	2.0		&	$6.4\times10^{34}$	&	18			&	﻿\cite{2016ApJ...826..142C,2018ApJ...853...27M} \\
V339~Del	&	Gaia Parallax: $0.381\pm0.361$~mas	&	2.1		&	$6.7\times10^{34}$	&	19			&	\cite{2014Sci...345..554A,2018MNRAS.481.3033S} \\
Nova Ret 2020 & Gaia Parallax: $0.316\pm0.046$~mas & 2.7 & $1.9\times10^{35}$ & 53 & \cite{2020ATel13868....1L,2018AJ....156...58B} \\
V407~Cyg	&	 Secondary star	classification &	2.7		&	$3.2\times10^{35}$	&	90			&	\cite{2014Sci...345..554A,1990MNRAS.242..653M} \\
V5855~Sgr	&	Magnitude on day 15 after peak	&	4.5		&	$7.1\times10^{35}$	&	201			&	\cite{2019ApJ...872...86N} \\
V5856~Sgr	&	Maximum magnitude-rate of decline				&	4.2		&	$8.2\times10^{35}$	&	232			&	\cite{2017NatAs...1..697L} \\
V906~Car	&	Reddening study			&	4.0		&	$1.7\times10^{36}$	&	490			&	\cite{2020NatAs.tmp...79A} \\
﻿V1324~Sco	&	Reddening study			&	$>6.5$	&	$>1.8\times10^{36}$	&	$>507$			&	﻿\cite{2014Sci...345..554A,2015ApJ...809..160F} \\
V392~Per	&	Gaia Parallax: $0.257\pm0.052$~mas	&	4.2		&	$3.8\times10^{36}$	&	1080			&	\cite{2018ATel11590....1L,2018MNRAS.481.3033S} \\
\hline
\label{tab:Lg}
\end{tabular}
\end{table*}

Since the Galactic foreground extinction should be low as mentioned in the previous section, the relatively high reddening observed in \src\ implies a high intrinsic extinction that is possibly related to the warm dust gas detected in the pre-nova WISE observations.

At $d=560$~pc, the marginally-detected $\gamma$-ray counterpart would have a very low luminosity of $L_\gamma\approx4\times10^{33}$\lum\ compared to other LAT-detected novae that typically radiate at $L_\gamma=10^{34-36}$\lum\ (Table \ref{tab:Lg}). Among all the $\gamma$-ray novae with estimated distances in the literature, V5668~Sgr is the least luminous $\gamma$-ray nova before \src, but its luminosity is still nearly an order of magnitude higher than that of \src. Even if $d=1.05$~kpc is adopted for \src, the luminosity of V5668~Sgr is still 2.5 times higher. The sub-luminous $\gamma$-ray emission of \src\ (and perhaps V5668~Sgr also) could be a hint of a slightly different scenario from that of the other brighter $\gamma$-ray novae. We suspect that the DES that the nova ejecta collided with is neither earlier slower ejecta (e.g., ASASSN-16ma; \citealt{2017NatAs...1..697L}) nor a pre-existing red-giant wind (e.g., V407 Cygni; \citealt{2010Sci...329..817A}), but the warm dust cloud surrounding the binary observed by WISE, of which the mass density is presumed to be low to produce the low $\gamma$-ray flux. 
The cloud could be a circumbinary disk, which was theoretically predicted in CVs \citep{2001ApJ...548..900S,2001ApJ...561..329T}.
Interestingly, there have been hints of pre-explosion winds in some other shock-powered transient classes, e.g., luminous red novae and Type IIn supernovae \citep{2017MNRAS.471.3200M}, which might be related to the origin of the circumbinary disk.
While the physical properties of the disk are mostly unclear, the hydrogen column density (\nh) of \src\ is around $\sim100$ times lower than that of V5855~Sgr and V906~Car (see \S\ref{sec:x2g}), likely indicating the low density of the dust cloud. Indeed, the \nh\ ratio is in the same order as the observed $\gamma$-ray luminosity ratio between \src\ and the more luminous ones. The shocks that only plow through such a low-density medium are likely non-radiative. 
While the non-radiative shocks could still produce optical emission, the efficiency would be much lower than that of the radiative ones observed in, e.g., ASASSN-16ma \citep{2017NatAs...1..697L} and V906~Car (\citealt{2020NatAs.tmp...79A}; see also \citealt{2014MNRAS.442..713M}). This makes the missing correlation between the $\gamma$-ray and optical light curves reasonable. We suspect that the $\gamma$-ray spectrum of \src\ is also different from those originating from radiative shocks (see the ``universal'' $\gamma$-ray spectral models for novae proposed by \citealt{2018A&A...609A.120F}). However, the LAT photon statistics of \src\ are just too poor for a meaningful comparison.

\section{Summary}
\src\ is a weak $\gamma$-ray nova with a poorly constrained \textit{Gaia} distance of 380--1050~pc (i.e., with an astrometric excess noise of $\epsilon_i=3.2$~mas). This makes \src\ potentially the least luminous $\gamma$-ray nova known to date. While the nearby \textit{Gaia} distance is consistent with that inferred from the supersoft X-ray emission as well as the progenitor's observations, the peak absolute magnitude of the nova is rather dim and this likely suggests a longer distance. Future \textit{Gaia} DR3 data may be able to clarify the issue. 

No correlation can be found between the optical and $\gamma$-ray light curves taken by ANS and the \textit{Fermi}-LAT, respectively. This could be due to the low S/N of the LAT data, or more likely, that the shock emission is not sufficiently strong to dominate in the optical band. In addition, the relatively low X-ray absorption during the $\gamma$-ray active phase (i.e., \nh~$\sim10^{22}$\cm) likely indicates that the nova ejecta was interacting with a low-density medium, possibly a circumbinary disk that was detected as the IR excess in the pre-nova observations. This provides a possible explanation for the low $\gamma$-ray luminosity observed, if it is a nearby nova. 

\begin{acknowledgements}

KLL is supported by the Ministry of Science and Technology of the Republic of China (Taiwan) through grants 108-2112-M-007-025-MY3 and 109-2636-M-006-017, and he is a Yushan (Young) Scholar of the Ministry of Education of the Republic of China (Taiwan).
LC acknowledges support from NASA/Fermi grant 80NSSC18K1746 and NSF grant AST-1751874.
J.S. acknowledges support from the Packard Foundation.
This work used high-performance computing facilities operated by the Center for Informatics and Computation in Astronomy (CICA) at National Tsing Hua University. This equipment was funded by the Ministry of Education of Taiwan, the Ministry of Science and Technology of Taiwan, and National Tsing Hua University.
This work has made use of data from the European Space Agency (ESA) mission
{\it Gaia} (\url{https://www.cosmos.esa.int/gaia}), processed by the {\it Gaia}
Data Processing and Analysis Consortium (DPAC,
\url{https://www.cosmos.esa.int/web/gaia/dpac/consortium}). Funding for the DPAC
has been provided by national institutions, in particular the institutions
participating in the {\it Gaia} Multilateral Agreement.
This publication makes use of data products from the Two Micron All Sky Survey, which is a joint project of the University of Massachusetts and the Infrared Processing and Analysis Center/California Institute of Technology, funded by the National Aeronautics and Space Administration and the National Science Foundation.
This publication makes use of data products from the Wide-field Infrared Survey Explorer, which is a joint project of the University of California, Los Angeles, and the Jet Propulsion Laboratory/California Institute of Technology, funded by the National Aeronautics and Space Administration.

\end{acknowledgements}
\textit{Facilities}: \facility{Fermi, Swift}

\bibliography{17mt}
\end{document}